\documentclass{emulateapj}

\usepackage{psfig,natbib,amsmath}

\def\sn{SN~2006gy}
\def\kms{km~s$^{-1}$}

\def\arcmin{\hbox{$^\prime$}}
\def\arcsec{\hbox{$^{\prime\prime}$}}
\def\msun{M$_\odot$}
\def\lsun{L$_\odot$}

\def\jhk{{\em J}, {\em H}, and {\em K$_s$}}
\def\etal{et al.\ }

\shorttitle{Echoes from \sn}

\shortauthors{Miller et al.}

\begin{document}

\title{New Observations of the Very Luminous Supernova 2006gy: 
Evidence for Echoes}

\def\berk{1}
\def\sloan{2}
\def\uco{3}
\def\ucsc{4}
\def\spitzer{5}

\author{A. A. Miller\altaffilmark{\berk}, 
N. Smith\altaffilmark{\berk}, 
W. Li\altaffilmark{\berk},
J. S. Bloom\altaffilmark{\berk,\sloan}, 
R. Chornock\altaffilmark{\berk},   
A. V. Filippenko\altaffilmark{\berk}, and
J. X. Prochaska\altaffilmark{\uco,\ucsc}
}

\altaffiltext{\berk}{Department of Astronomy, University of California, Berkeley, CA 94720-3411.}

\altaffiltext{\sloan}{Sloan Research Fellow.}

\altaffiltext{\uco}{University of California Observatories/Lick Observatory, University of California, Santa Cruz, CA 95064.}

\altaffiltext{\ucsc}{Department of Astronomy and Astrophysics, University of California, Santa Cruz, CA 95064.}

\begin{abstract}

Supernova (SN) 2006gy was a hydrogen-rich core-collapse SN that remains one 
of the most luminous optical supernovae ever observed. The total energy 
budget ($>$ 2$\times$10$^{51}$ erg radiated in the optical alone) poses 
many challenges for standard SN theory.  
We present new ground-based near-infrared (NIR) observations of \sn, as well 
as a single epoch of {\it Hubble Space Telescope (HST)} imaging obtained more 
than two years after the explosion. Our NIR data taken around peak optical 
emission show an evolution that is largely consistent with a cooling 
blackbody, with tentative evidence for a growing NIR excess starting 
around day $\sim$130. Our late-time Keck adaptive optics NIR
image, taken on day 723, shows little change from previous NIR observations 
taken around day 400. Furthermore, the optical {\it HST} observations show 
a reduced decline rate after day 400, and the SN is bluer on day 810 
than it was at peak. This late-time decline is inconsistent with $^{56}$Co 
decay, and thus is problematic for the various pair-instability SN models 
used to explain the nature of \sn. The slow decline of the NIR emission can be 
explained with a light echo, and we confirm that the 
late-time NIR excess is the result of a massive ($\ga$10 \msun) dusty 
shell heated by the SN peak luminosity. The late-time optical observations 
require the existence of a scattered light echo, which may be generated 
by the same dust that contributes to the NIR echo. Both the NIR and optical 
echoes originate in the proximity of the progenitor, $\sim$10$^{18}$ cm 
for the NIR echo and $\la$10--40 pc for the optical echo, which provides 
further evidence that the progenitor of \sn\ was a very massive star.

\end{abstract}


\section{Introduction}

At the time of discovery, supernova (SN) 2006gy was the most luminous 
SN ever found \citep{ofek06gy, smith07-2006gy}. \sn\ 
generated a great deal of interest; in addition to being $\sim$100 
times more luminous than a typical Type II (hydrogen-rich, core-collapse) 
SN at peak, it exhibited a 
long rise time ($\sim$70 day) and slow decline, leading to speculation 
that it may have been the first observed example of a pair-instability SN
\citep[PISN;][]{ofek06gy, smith07-2006gy}\footnote{Currently, there is 
much stronger evidence that SN~2007bi is the first observed example of 
a PISN \citep{gal-yam07bi}.} or a pulsational pair-instability 
SN \citep{woosley07}. 

\sn\ was classified as a Type IIn SN (see \citealt{schlegel1990} for a 
definition of the Type IIn subclass and \citealt{filippenko1997} for a review 
of its spectral properties) based on the relatively narrow emission 
features present in the early-time SN spectrum. Some Type IIn supernovae 
(SNe~IIn) are known to be 
overluminous relative to their typical SN~II counterparts: $M_R \approx -15.8$ 
mag for Type~II-P with a 1$\sigma$ scatter of 1.1 mag  
\citep{li-ratesLF}, whereas \sn\ reached $M_R \approx -21.7$ mag. The 
enhanced luminosity of some SNe IIn is probably due to the collision of 
fast-moving SN ejecta with a dense, 
and possibly clumpy, circumstellar medium (CSM; e.g., \citealt{chugai1994}). 
In a companion paper \citep{smith10-2006gy}, a detailed spectroscopic 
comparison of \sn\ is made to other SNe~IIn. \sn\ is unique within 
the SN~IIn subclass, however, because typical interaction models cannot 
explain its early-time behavior, suggesting the need 
for alternative models for this particular object 
\citep{smith07-2006gy,woosley07,nomoto07,smith10-2006gy}.

Pair-instability SNe \citep{barkat67, rs1967, bond84} are expected to occur in 
very massive, low-metallicity stars, such as those that may have been 
present in the metal-free environment of the very early universe (e.g., 
\citealt{abel2000}). The detection of a pair-instability SN in the 
comparatively local universe, then,  
could potentially reveal a great deal about the first generation of stars.
The light curves of pair-instability SNe are expected to exhibit a relatively 
slow rise, followed by a broad turnover after the peak, and a peak luminosity 
that is considerably larger than that of typical SNe \citep{scannapieco2005}. 
Qualitatively, each of these characteristics matches those observed for 
\sn.

Other more luminous SNe have been announced since the discovery of \sn: SNe 
2005ap \citep{quimby05ap} and 2008es \citep{miller08es, gezari08es}. This
suggests that while these events are rare, there may be a new subclass 
of very luminous supernovae (VLSNe). The peak 
luminosity and photometric evolution of both SNe 2005ap and 2008es are 
difficult to explain via the pair-instability model \citep{quimby05ap, 
miller08es}, implying that peak luminosities $\ga$ few $\times 10^{44}$ 
erg s$^{-1}$ are possible without a pair-instability explosion.  
There is a wide diversity of alternative models that have been developed 
to explain the early-time observations of \sn. \citet{smith+mccray07} argue 
that the peak luminosity and light-curve evolution can be explained via
the thermalization of shock energy deposited into a massive ($\sim$10 \msun), 
optically thick shell. Based on a model of the light curve 
near peak, \citet{agnoletto09} suggest that the combination of  
ejecta colliding with dense clumps in the CSM and $\sim$1--3 \msun\ 
of $^{56}$Ni are responsible for the early-time luminosity of \sn.
\citet{nomoto07} are unable to match the light curve with a standard 
pair-instability 
SN model; however, they can reasonably reproduce the ($\la$ 400 day) light 
curve of \sn\ using a pair-instability model where they artificially reduce 
the ejecta mass, such that the radioactive heating is less than 100\% 
efficient. The reduction in ejecta mass should lead to a reduction in 
$^{56}$Ni production, as noted by \citet{nomoto07}; thus, their model 
with an artificially reduced ejecta mass may not be self consistent.
\citet{woosley07} use a model of a pulsational pair instability 
within a massive star to explain the light curve of \sn.

Some of the models make predictions for the late-time 
behavior of \sn. Those with a large yield of $^{56}$Ni 
\citep{nomoto07, smith07-2006gy} would expect a decline in bolometric 
luminosity at the rate of $^{56}$Co decay, 0.98 mag (100 day)$^{-1}$, or 
faster if the radioactive-decay energy is not converted to optical emission 
with 100\% efficiency. 
The shell-shock model \citep{smith+mccray07} predicts 
a rapid decline after $\sim$200 day, though the authors note that this 
decline may be offset by the production of a large amount of $^{56}$Ni 
or continued CSM interaction. The 
pulsational pair-instability model \citep{woosley07} predicts another SN
explosion at the location of \sn\ about 9 years after the initial outburst 
from \sn. 

More than a year after the explosion, \citet{smith08-2006gy} 
detected \sn\ in the near infrared (NIR) at a luminosity comparable to 
that of the peak luminosity of most SNe~II. This property had not 
been predicted by any of the models. When coupled with the 
lack of a detection in the radio and X-rays, this led \citet{smith08-2006gy} 
to conclude that the luminosity could not be powered by the continued 
interaction of the SN ejecta with CSM. \citet{kawabata09} draw similar 
conclusions based on their detection of weak H$\alpha$ emission at a 
comparable epoch. \citet{smith08-2006gy} conclude that only two 
possibilities are able to explain the late-time observations of \sn: (i) 
the explosion produced $\ga$2.5 \msun\ of $^{56}$Ni, which is only 
theoretically expected for PISNe (e.g., \citealt{scannapieco2005}), 
which was heating dust and consequently generating 
the large NIR excess, or (ii) a massive ($\sim$5--10 \msun), dusty shell, 
located $\sim$1 light year from the site of the SN, was being heated by the 
radiation produced at peak, and reradiating that energy as a NIR echo 
\citep{dwek1983}. For the first case, if the luminosity were powered by 
radioactive decay, then future observations should indicate a continued 
decline at the rate of $^{56}$Co decay. A dust echo, on the other 
hand, would result in a NIR light curve that stays roughly constant for
$\sim$1--2 yr before exhibiting a rapid decline.

Evolved massive stars, such as red supergiants 
and luminous blue variables (LBVs), are often observed to 
have massive dust shells, so if these stars explode as SNe~IIn
one might expect a late-time IR echo. Many SNe~IIn have been observed to 
exhibit a late-time NIR excess \citep{gerardy2002} which in some cases 
lasted $>$ 1 yr. This excess has been attributed 
to NIR echoes \citep{gerardy2002}, though we note that the formation 
of new dust has been argued specifically for SN 1998S  
\citep{pozzo2004} and SN 1995N \citep{fox2009}. Dusty regions near the SN 
should also lead to ultraviolet (UV) and
optical scattered-light echoes \citep{chevalier1986}; thus, dust is 
capable of providing significant optical and NIR emission at late times.
The optical decline of SNe~IIn at late times is very heterogeneous 
\citep{WLi2002}, and therefore caution must be applied when determining 
the source of any late-time emission.

In this paper we present new NIR observations of \sn, taken around the peak of 
optical emission, as well as optical and NIR observations obtained more 
than two years after \sn\ exploded. Section 2 describes the observations 
and data reduction. We discuss the results in $\S$ 3, and in $\S$ 4 we offer 
some conclusions. Throughout this paper we assume that the distance to NGC 1260
(the host galaxy of \sn) is 73.1 Mpc, and following \citet{smith07-2006gy} 
we adopt $E(B-V) = 0.54$ mag as the reddening toward \sn\ within the host 
galaxy, while Galactic extinction accounts for $E(B-V) = 0.18$ mag, leading 
to a total color excess toward \sn\ of $E(B-V) = 0.72$ mag. 
All spectral energy 
distributions (SEDs) have been corrected for this color excess assuming 
$R_V = A_V/E(B-V) = 3.1$ using the reddening law of \citet{ccm1989}.

\section{Observations}

NIR observations of \sn\ were obtained simultaneously in \jhk\ 
with the Peters Automated Infrared Imaging Telescope (PAIRITEL; 
\citealt{bloom06}) starting on 2006 October 13 UT\footnote{All dates in 
this paper are UT unless otherwise noted.} (54 days after 
explosion\footnote{Following \citet{smith07-2006gy}, we adopt 2006 August 20 
as the explosion date for \sn.}). PAIRITEL is a 1.3-m robotic telescope, 
located on Mt. Hopkins, AZ, which obtained images over the next
24 months for 0.5--1 hr at each of the 124 epochs through normal 
queue-scheduled operations. All images were processed via an 
automated pipeline \citep{bloom06}.

\begin{figure*}
\plotone{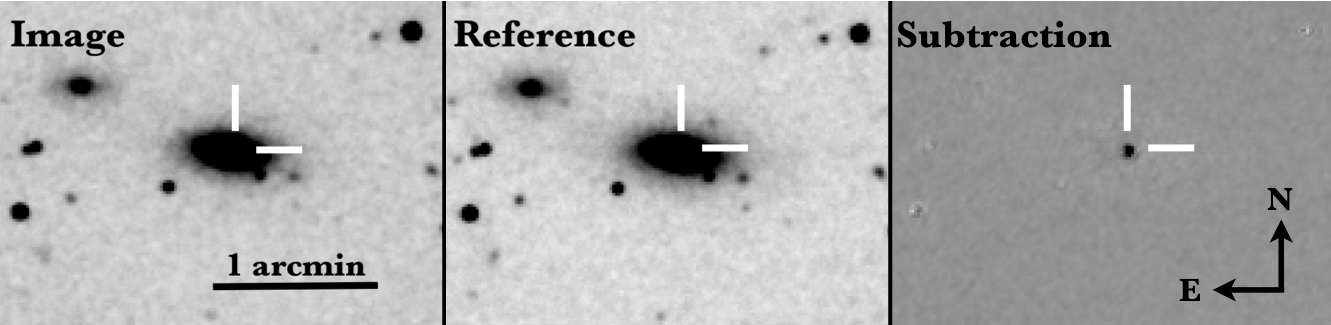}
\caption{Example subtraction of PAIRITEL NIR images of \sn, which have 
all been registered to the same coordinate frame. Each image is 
$\sim$2\arcmin\ $\times$ 3\arcmin\ in size. {\it Left}: PAIRITEL 
$K_s$-band image taken on 2006 Nov. 15. {\it Middle}: PAIRITEL $K_s$-band 
image taken on 2007 Jan. 01. {\it Right}: Image (left panel) minus reference 
(middle panel) subtraction image. The image subtraction was performed with 
HOTPANTS, and \sn\ is 
clearly visible in the difference image as a bright point source near the 
galaxy nucleus.} 
\label{ptel-subtract}
\end{figure*}

Analysis of \sn\ has proved challenging because the SN is located very 
close to the nucleus of NGC 1260 (separation $\sim 1\arcsec$; 
\citealt{ofek06gy, smith07-2006gy}). PAIRITEL has a large native scale of 
2$\arcsec$ pixel$^{-1}$, precluding spatial resolution of the SN from the 
nucleus. This necessitates image differencing to obtain the light curve
of the SN. Image subtraction was performed with 
HOTPANTS\footnote{http://www.astro.washington.edu/users/becker/hotpants.html .}, 
and the flux in each difference pair was determined via aperture photometry 
at the location of the SN. An example subtraction, which clearly shows flux 
from \sn\ after the reference image has been subtracted, is shown in 
Figure~\ref{ptel-subtract}.

Despite observations extending more than two 
years past the date of discovery, \sn\ has not faded beyond the point of
detectability with PAIRITEL. Consequently, all \jhk\ images 
of \sn\ contain some flux from the SN. We thus adopted the ``NN2 method'' 
of \citet{barris2005-NN2} to determine the relative NIR flux changes of 
the SN.  The NN2 method treats all images equally and does not require a 
template image with no light from the source of interest, \sn. It
uses the subtraction of all $N(N-1)$ pairs of images 
to mitigate against possible errors associated with the use of a single 
reference template image. The downside to the NN2 method is that it only 
produces the differential flux between each of the $N$ epochs of observations. 
To convert these flux differences into magnitudes requires an absolute 
calibration, which must be obtained independently of the results from the 
NN2 method. Uncertainties in the individual subtractions were 
estimated by measuring the scatter in fake SNe inserted at locations having a 
surface brightness similar to that at \sn.

\begin{figure}
\epsscale{1.15}
\plotone{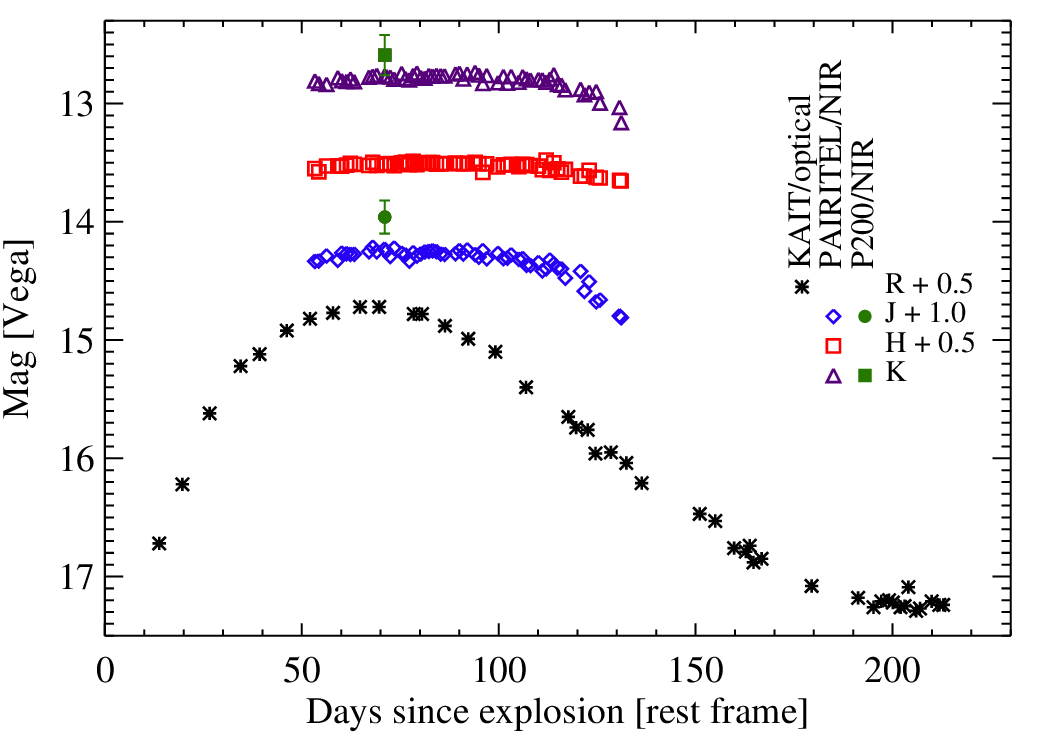}
\caption{Early-time photometric evolution of \sn. Unfiltered KAIT (roughly
$R$ band) observations are taken from \citet{smith07-2006gy}. PAIRITEL \jhk\ 
observations are from this work. We also show the Palomar AO 
photometry from \citet{ofek06gy}. The data have not been corrected for 
extinction in the host or the Galaxy. The NIR evolution is remarkably 
flat, with the $R-K$ color increasing steadily after day $\sim$70. This 
behavior is consistent with a cooling blackbody (see text).} 
\label{early-lc}
\end{figure}

Early-time NIR observations of \sn\ (defined here as those made before 
\sn\ passed behind the Sun during 2007) show a remarkably flat light curve, 
as seen in Figure~\ref{early-lc}. To transform the relative-flux 
differences from the NN2 method to an absolute scale, we 
subtracted the archival Two Micron All Sky Survey (2MASS;
\citealt{skrutskie-2mass}) image from each of the images obtained on 
2006 Nov. 13.24, 14.25, 15.38, and 16.37 to determine the 
\jhk\ magnitudes of the SN on these dates. The mean SN flux was 
then used to transform the relative flux from the NN2 method to 
the absolute flux of the 2MASS system \citep{cohen2003}. PAIRITEL uses 
the old 2MASS camera and telescope; hence, we do not expect any 
large systematic effects in the 2MASS subtractions. The final 
calibrated \jhk\ photometry is summarized in Table~\ref{ptel-data}.

On 2006 Nov. 01, \citet{ofek06gy} obtained adaptive optics (AO) 
images in the $J$ and $K_s$ bands with the Palomar Hale (5~m) 
telescope, clearly resolving \sn\ from 
the host-galaxy nucleus, and measured its flux. In the $K_s$ band 
our calibration and the \citet{ofek06gy} measurement agree to within 
1$\sigma$, while the agreement in the
$J$ band is somewhat worse ($\sim$2$\sigma$). \citet{ofek06gy} had only a 
single 2MASS star within the field of view of their AO images, whereas 
$>$100 2MASS stars were used to calibrate the PAIRITEL images; 
when coupled with the difficulty associated with photometry
of AO images, this may explain the differences between the two measurements. 
We note that were we to adopt 
the \citet{ofek06gy} measurements as our calibration, there would be
an overall systematic shift of our $J$ and $K_s$ light curves to 
brighter values, which in turn would lead to worse agreement between the 
NIR data and early-time optical spectra (see Figure~\ref{early-SED}). This 
suggests that our calibration method is sufficient. We note that 
the uncertainties in our photometry are dominated by the uncertainty in 
the calibration, which is $\sim$0.03 mag in $J$, $\sim$0.06 mag in 
$H$, and $\sim$0.04 mag in $K_s$. This uncertainty is 
the same for all epochs, so a change in 
the calibration would lead to a systematic shift of the entire light curve. 

\begin{table}
\begin{minipage}{80mm}
	\centering
\caption{PAIRITEL Observations of \sn}
\begin{tabular}{rccc}
\hline\hline
$t_{\rm mid}$\footnote{Midpoint between the first and last exposures in a single stacked image.}
 & $J$ mag\footnote{Observed value; not corrected for Galactic extinction.}
 & $H$ mag$^{\rm b}$ & $K_s$ mag$^{\rm b}$ \\ 
(MJD) & (Vega) & (Vega) & (Vega) \\
\hline
54021.29 &  13.33 $\pm$  0.04 &  13.05 $\pm$  0.06 &  12.81 $\pm$  0.05 \\
54022.28 &  13.33 $\pm$  0.04 &  13.08 $\pm$  0.09 &  12.83 $\pm$  0.05 \\
54024.30 &  13.29 $\pm$  0.04 &  13.03 $\pm$  0.06 &  12.84 $\pm$  0.06 \\
54027.22 &  13.32 $\pm$  0.04 &  13.03 $\pm$  0.06 &  12.78 $\pm$  0.04 \\
54028.23 &  13.26 $\pm$  0.04 &  13.03 $\pm$  0.06 &  12.81 $\pm$  0.05 \\
54029.46 &  13.27 $\pm$  0.03 &  13.02 $\pm$  0.06 &  12.81 $\pm$  0.06 \\
54030.47 &  13.27 $\pm$  0.04 &  13.01 $\pm$  0.06 &  12.79 $\pm$  0.04 \\
54031.45 &  13.27 $\pm$  0.04 &  13.01 $\pm$  0.06 &  12.81 $\pm$  0.05 \\
54035.26 &  13.25 $\pm$  0.04 &  13.02 $\pm$  0.06 &  12.78 $\pm$  0.04 \\
54036.25 &  13.21 $\pm$  0.05 &  13.00 $\pm$  0.06 &  12.77 $\pm$  0.04 \\
54037.28 &  13.25 $\pm$  0.03 &  13.02 $\pm$  0.06 &  12.76 $\pm$  0.04 \\
54039.28 &  13.23 $\pm$  0.04 &  13.01 $\pm$  0.06 &  12.77 $\pm$  0.04 \\
54040.29 &  13.25 $\pm$  0.04 &  13.01 $\pm$  0.06 &  12.77 $\pm$  0.05 \\
54041.31 &  13.29 $\pm$  0.04 &  13.01 $\pm$  0.06 &  12.78 $\pm$  0.04 \\
54042.32 &  13.22 $\pm$  0.05 &  13.03 $\pm$  0.08 &  12.79 $\pm$  0.09 \\
54044.27 &  13.26 $\pm$  0.04 &  13.00 $\pm$  0.06 &  12.74 $\pm$  0.04 \\
54045.27 &  13.28 $\pm$  0.05 &  13.00 $\pm$  0.06 &  12.79 $\pm$  0.05 \\
54046.25 &  13.33 $\pm$  0.04 &  13.02 $\pm$  0.07 &  12.80 $\pm$  0.05 \\
54047.25 &  13.26 $\pm$  0.03 &  12.99 $\pm$  0.06 &  12.76 $\pm$  0.05 \\
54048.23 &  13.29 $\pm$  0.04 &  13.02 $\pm$  0.06 &  12.74 $\pm$  0.05 \\
54049.23 &  13.27 $\pm$  0.04 &  13.01 $\pm$  0.06 &  12.78 $\pm$  0.05 \\
54050.27 &  13.25 $\pm$  0.03 &  13.00 $\pm$  0.06 &  12.78 $\pm$  0.04 \\
54051.31 &  13.25 $\pm$  0.04 &  13.00 $\pm$  0.06 &  12.76 $\pm$  0.04 \\
54052.25 &  13.24 $\pm$  0.04 &  13.00 $\pm$  0.06 &  12.77 $\pm$  0.04 \\
54053.26 &  13.25 $\pm$  0.03 &  13.01 $\pm$  0.06 &  12.76 $\pm$  0.04 \\
54054.38 &  13.27 $\pm$  0.04 &  13.01 $\pm$  0.06 &  12.76 $\pm$  0.04 \\
54055.38 &  13.28 $\pm$  0.04 &  13.01 $\pm$  0.06 &  12.77 $\pm$  0.04 \\
54058.18 &  13.27 $\pm$  0.03 &  13.00 $\pm$  0.06 &  12.75 $\pm$  0.04 \\
54059.19 &  13.24 $\pm$  0.04 &  13.01 $\pm$  0.06 &  12.74 $\pm$  0.05 \\
54060.22 &  13.27 $\pm$  0.03 &  13.01 $\pm$  0.06 &  12.79 $\pm$  0.04 \\
54061.21 &  13.24 $\pm$  0.04 &  13.01 $\pm$  0.06 &  12.75 $\pm$  0.05 \\
54063.24 &  13.28 $\pm$  0.04 &  13.00 $\pm$  0.06 &  12.74 $\pm$  0.05 \\
54064.24 &  13.30 $\pm$  0.04 &  13.01 $\pm$  0.06 &  12.76 $\pm$  0.04 \\
54065.24 &  13.24 $\pm$  0.07 &  13.08 $\pm$  0.11 &  12.83 $\pm$  0.19 \\
54066.25 &  13.31 $\pm$  0.04 &  13.01 $\pm$  0.06 &  12.76 $\pm$  0.05 \\
54069.15 &  13.27 $\pm$  0.04 &  13.04 $\pm$  0.06 &  12.82 $\pm$  0.05 \\
54070.13 &  13.31 $\pm$  0.04 &  13.02 $\pm$  0.06 &  12.77 $\pm$  0.04 \\
54071.15 &  13.31 $\pm$  0.05 &  13.02 $\pm$  0.06 &  12.83 $\pm$  0.05 \\
54072.13 &  13.28 $\pm$  0.04 &  13.01 $\pm$  0.06 &  12.77 $\pm$  0.04 \\
54074.09 &  13.31 $\pm$  0.04 &  13.03 $\pm$  0.06 &  12.82 $\pm$  0.05 \\
54075.13 &  13.31 $\pm$  0.04 &  13.01 $\pm$  0.06 &  12.77 $\pm$  0.05 \\
54076.12 &  13.37 $\pm$  0.04 &  13.02 $\pm$  0.06 &  12.79 $\pm$  0.04 \\
54077.12 &  13.37 $\pm$  0.04 &  13.02 $\pm$  0.06 &  12.80 $\pm$  0.04 \\
54079.18 &  13.34 $\pm$  0.05 &  13.03 $\pm$  0.06 &  12.80 $\pm$  0.05 \\
54080.18 &  13.41 $\pm$  0.05 &  13.06 $\pm$  0.06 &  12.80 $\pm$  0.04 \\
54081.17 &  13.40 $\pm$  0.04 &  12.98 $\pm$  0.06 &  12.82 $\pm$  0.05 \\
54082.15 &  13.32 $\pm$  0.07 &  13.07 $\pm$  0.07 &  12.79 $\pm$  0.05 \\
54083.16 &  13.36 $\pm$  0.09 &  13.00 $\pm$  0.06 &  12.75 $\pm$  0.05 \\
54084.14 &  13.39 $\pm$  0.05 &  13.05 $\pm$  0.06 &  12.84 $\pm$  0.08 \\
54085.13 &  13.40 $\pm$  0.05 &  13.08 $\pm$  0.07 &  12.84 $\pm$  0.05 \\
54086.13 &  13.48 $\pm$  0.06 &  13.06 $\pm$  0.07 &  12.88 $\pm$  0.06 \\
54090.08 &  13.42 $\pm$  0.04 &  13.11 $\pm$  0.07 &  12.88 $\pm$  0.05 \\
54091.08 &  13.59 $\pm$  0.09 &  13.11 $\pm$  0.07 &  12.93 $\pm$  0.05 \\
54092.29 &  13.51 $\pm$  0.11 &  13.06 $\pm$  0.06 &  12.90 $\pm$  0.07 \\
54094.12 &  13.68 $\pm$  0.11 &  13.12 $\pm$  0.07 &  12.90 $\pm$  0.05 \\
54095.12 &  13.66 $\pm$  0.11 &  13.13 $\pm$  0.07 &  13.00 $\pm$  0.16 \\
54100.14 &  13.79 $\pm$  0.08 &  13.15 $\pm$  0.07 &  13.03 $\pm$  0.13 \\
54101.15 &  13.81 $\pm$  0.07 &  13.16 $\pm$  0.07 &  13.16 $\pm$  0.14 \\ 
\hline
\end{tabular}
\label{ptel-data}
\end{minipage}
\end{table}

As shown by \citet{smith08-2006gy}, 
and subsequently confirmed by \citet{agnoletto09}, the NIR evolution of \sn\ is very 
slow at late times. Given the relatively small change in flux, and the reduced 
signal-to-noise ratio
following the fading of the SN, we were unable to recover reliable flux 
measurements from PAIRITEL data taken after 2007 Sep. 
Furthermore, unlike the case at early times, the 
SN had faded below the 2MASS detection limit, which means that the late-time 
subtractions relative to the 2MASS template image do not yield meaningful 
results despite the fact that at $K_s \approx 15$ the SN is well above the 
PAIRITEL detection limit. In principle, if deep PAIRITEL images are obtained 
after \sn\ fades well beyond the detection limit, it should be 
possible to recover the late-time NIR light curve using template images 
that contain little or no light from the SN.

We also observed \sn\ on 2008 Aug. 25 with the Near-Infrared Camera 2 
(NIRC2) using the laser guide star (LGS) AO system \citep{wizinowich2006} 
on the 10-m Keck II telescope in Hawaii. We have rereduced the LGS AO images 
presented by \citet{smith08-2006gy} from days 398 and 461\footnote{Note that 
\citet{smith08-2006gy} refer to these epochs as day 405 and 468, which is the 
elapsed time in the observed frame. All epochs in the present work are 
labelled in terms of the elapsed time in the rest frame.} and derive revised 
values for the $K'$-band magnitudes of \sn, as summarized in 
Table~\ref{tbl-KeckAO}.\footnote{Late-time Keck AO observations were all made 
in the $K'$ band. The uncertainty associated with the transformation between 
$K'$ and $K_s$ is small compared to the absolute calibration 
uncertainty; hence, for the late-time AO images $K' \approx K_s$.} The 
uncertainty on these measurements is large ($\ga 0.17$ mag), and is dominated 
by the uncertainty in the single calibration star within the field of view. 
We also present the first measurement of the $H$-band flux from the AO images 
taken on 2007 Dec. 2. To obtain this $H$-band 
measurement, despite a lack of 2MASS stars in the field, we measured the 
$H - K'$ color of \sn\ relative to the $H - K'$ color of the host galaxy, 
and calibrated this against the $H - K_s$ color of the galaxy in the archival 
2MASS images. The large uncertainty for this measurement reflects the 
accuracy with which we can determine the color of the galaxy from the 
2MASS images.

\begin{table}
\begin{minipage}{80mm}
	\centering
\caption{Keck AO Observations of \sn}
\begin{tabular}{lccc}
\hline\hline
date & epoch\footnote{Rest-frame days from the adopted explosion date, 2006 Aug.\ 20 \citep{smith07-2006gy}.} & Filter & mag\footnote{Observed value; not corrected for Galactic extinction.}
 \\ 
(UT) & (day) &  & (Vega) \\
\hline
2007 Sep. 29 & 398 & $K'$ & 14.91 $\pm$ 0.17 \\
2007 Dec. 02 & 461 & $H$ & 16.8 $\pm$ 0.3 \\
2007 Dec. 02 & 461 & $K'$ & 15.02 $\pm$ 0.17 \\
2008 Aug. 25 & 723 & $K'$ & 15.59 $\pm$ 0.21 \\
\hline
\end{tabular}

\label{tbl-KeckAO}
\end{minipage}
\end{table}

\begin{figure}
\epsscale{1.1}
\plotone{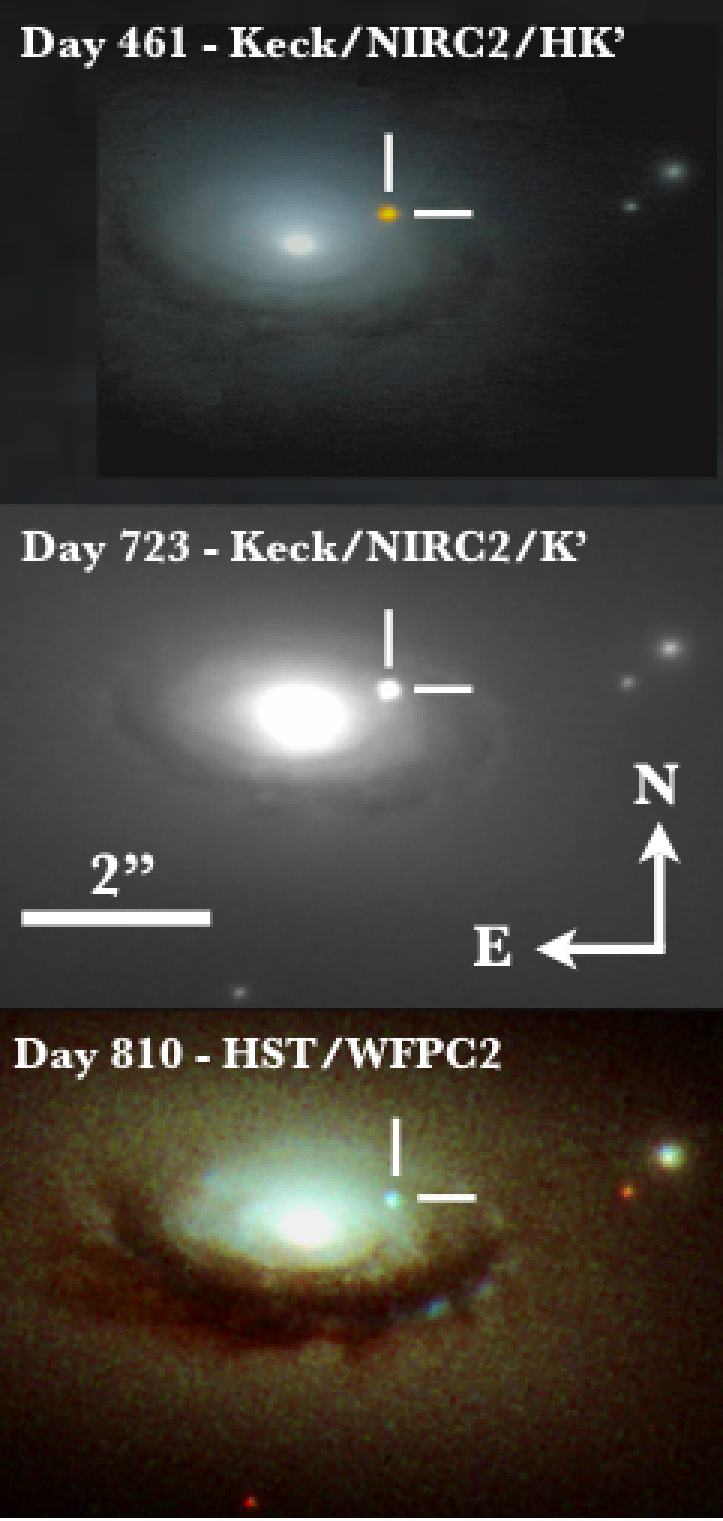}
\caption{Late-time optical and NIR images of \sn. {\it Top}:
Keck AO $H$ and $K'$ false-color image from \citet{smith08-2006gy}. 
The SN is clearly red in the NIR. Note that the field of view for this 
image is slightly smaller than that of the others. 
{\it Middle}: Keck AO $K'$ image of \sn\ taken on day 723. The SN is 
still clearly visible in the NIR, and has shown little change since
 observations taken around day 400. {\it Bottom}: 
{\it HST} WFPC2 false-color image including the four filters in which we 
detect \sn: $F450W$ (corresponding to blue), $F555W$, $F675W$, and 
$F814W$ (corresponding to red). The SN is clearly blue in the optical 
compared to the light from the surrounding stars. 
} 
\label{false-color}
\end{figure}

As part of a {\it Hubble Space Telescope (HST)} snapshot survey (GO-10877; 
PI Li), \sn\ was 
observed with the Wide Field Planetary Camera 2 (WFPC2; 
\citealt{holtzman1995a}) using the $F450W$, $F555W$, $F675W$, and $F814W$ 
filters on 2008 Nov. 22. The data were reduced in the standard fashion using 
{\bf multidrizzle} \citep{koekemoer2002}. NGC 1260 and SN 2006gy were 
located at the center of the PC chip of WFPC2, which has a native 
pixel scale of 0.0455$\arcsec$ pixel$^{-1}$. We follow the recipe of 
\citet{dolphin2000} to do charge-transfer efficiency correction and 
photometric reduction of the WFPC2 images. Our measurements of the 
\sn\ magnitudes are summarized in Table~\ref{tbl-HST}.
A false-color image of this detection is shown in Figure~\ref{false-color}.

\begin{table}
\begin{minipage}{80mm}
\caption{HST day 810 Observations of \sn\label{tbl-HST}}
\begin{tabular}{lcccc}
\hline\hline
Filter & $\lambda_{\rm cent}$\footnote{Central wavelength of the filter, based on WFPC2 calibrations presented in Table~8 of \citet{holtzman1995b}.} & $\Delta\lambda$\footnote{Filter width, based on WFPC2 calibrations presented in Table~8 of \citet{holtzman1995b}.} & mag\footnote{Observed value; not corrected for Galactic extinction.} & Flux\footnote{Observed flux using the photometric calibrations presented in Table~9 of \citet{holtzman1995b}.} \\ 
& (\AA) &  (\AA) & (Vega) & ($\mu$Jy) \\
\hline
$F450W$ & 4519 & 957 & 22.16 $\pm$ 0.03 & 5.97 $\pm$ 0.18 \\
$F555W$ & 5397 & 1226 & 21.51 $\pm$ 0.02 & 9.31 $\pm$ 0.19 \\
$F675W$ & 6697 & 866 & 21.07 $\pm$ 0.03 & 11.29 $\pm$ 0.34 \\
$F814W$ & 7924 & 1500 & 20.91 $\pm$ 0.04 & 10.80 $\pm$ 0.43 \\
\hline
\end{tabular}
\end{minipage}
\end{table}

\section{Discussion}

\subsection{Early-Time NIR Observations}

\begin{figure}
\epsscale{1.15}
\plotone{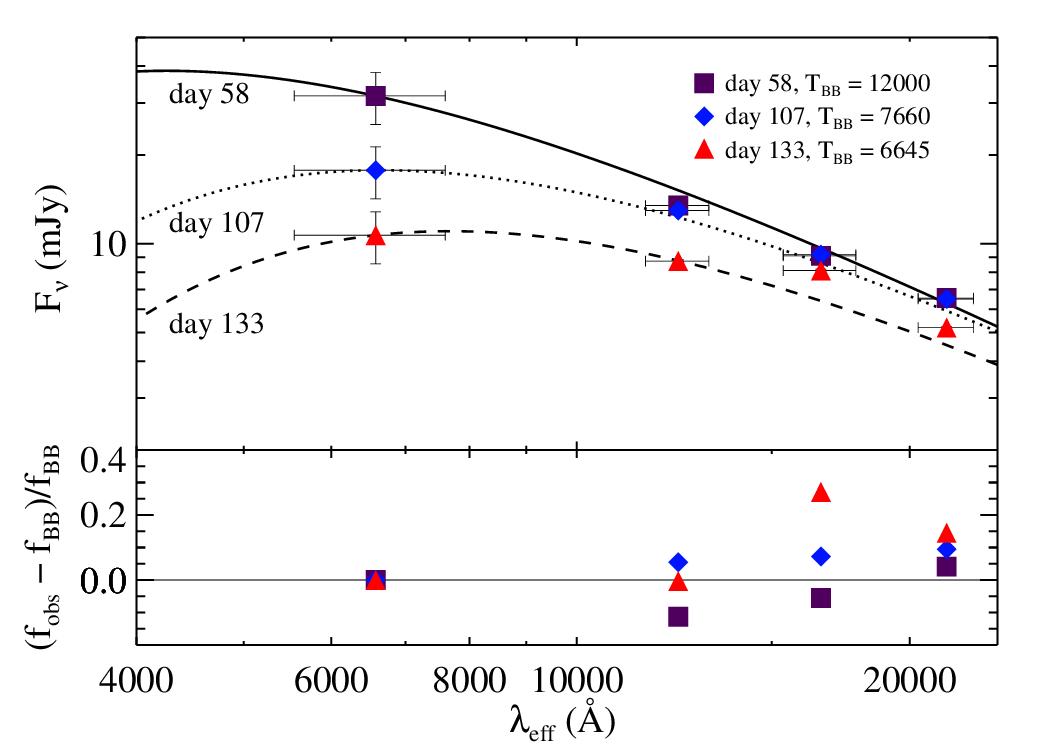}
\caption{SED evolution of \sn\ at early times showing a possible trend 
toward a growing IR excess above a cooling blackbody. {\it Top:} the solid, 
dotted, and 
dashed lines show a single-component blackbody spectrum, normalized to 
unfiltered observations from \citet{smith07-2006gy}, on days 58, 107, and 133, 
respectively. {\it Bottom:} the fractional excess emission in \jhk, relative 
to the 
blackbody model. For clarity we do not include the uncertainties in these 
measurements, which are large and dominated by the uncertainty in the 
transformation between unfiltered data and the $R$ band (see text). Each 
point is 
within $\sim$1$\sigma$ of showing no excess; however, the NIR excess does 
appear to grow with time.}
\label{early-SED}
\end{figure}

While the early-time evolution of \sn\ is remarkably flat in the NIR (see 
Figure~\ref{early-lc}), we find that these measurements are  
consistent with radiation from a cooling blackbody. From blackbody fits 
to optical spectra, \citet{smith10-2006gy} find that the temperature of 
\sn\ monotonically cools after day $\sim$50, roughly two weeks before optical 
maximum, until day $\sim$165, where the temperature levels off at 
$\sim$6300~K. In the top panel of Figure~\ref{early-SED} we show the 
evolution of the SED of \sn\ for three epochs (day 58, 107, 
and 133) during our early-time NIR observations. These three epochs were 
chosen as a representative sample covering the full range of our early NIR 
observations. The evolution of the SED is gradual; the three epochs shown in 
Figure~\ref{early-SED} are not more or less statistically significant, in 
terms of the observed NIR excess (see below), than other epochs from similar 
times. For each epoch we show the 
single-component blackbody spectrum, after adopting the temperature from 
\citet{smith10-2006gy} and normalizing the spectra to the photometric 
measurements from \citet{smith07-2006gy}. Their photometric 
observations come from a series of unfiltered observations taken with the 
Katzman Automatic Imaging Telescope (KAIT; \citealt{filippenkoLOSS}), which 
are best matched by the $R$ band \citep{Riess1999, WLi2003}. 
However, the scatter in the transformation between KAIT unfiltered and 
$R$ can be quite large \citep{mo-pc-2009}, so 
we adopt a 0.1 mag uncertainty for the calibration of the 
blackbody spectra. We also show the NIR flux during these epochs,
determined from the absolute calibration of the 2MASS system  
\citep{cohen2003}. This shows excellent agreement with an 
extrapolation of the early-time spectra of \sn\ \citep{smith10-2006gy}. 
Between day $\sim$55--135 roughly 2--4\% of 
the bolometric luminosity of \sn\ was emitted in the NIR.

In the lower panel of Figure~\ref{early-SED} we show the fractional excess 
of the photometric observations relative to the single-component blackbodies 
(note that by definition the $R$-band excess is set to zero). For clarity we 
do not show the uncertainties associated with the excess, but after 
accounting for the large uncertainty in the $R$-band calibration, each point 
is within $\sim$1$\sigma$ of zero excess. Nevertheless, there is an 
apparent trend that the excess is growing in the $H$ and $K_s$ bands 
as a function of time. The trend 
may be indicative of radiation from warm dust (see below), though we note that 
this effect would be small. If this trend toward a NIR excess is created 
by the same source as the late-time NIR excess, we would expect a $K_s$-band 
excess of $\sim$0.17 mag, which is comparable to the combined uncertainty 
from the blackbody calibration and photometric measurements.

\subsection{Late-Time NIR Observations}

With data obtained more than a year after explosion, 
\citet{smith08-2006gy} discovered a significant NIR excess from \sn. Our NIR
observations taken 723 days after explosion show that the $K'$-band flux from 
\sn\ has only faded by a factor of $\sim$2 over the course of the previous 
year. We cannot determine the total IR luminosity at late times, because our 
NIR data do not cover the peak of the SED. Furthermore, as first noted by 
\citet{smith08-2006gy}, the very red $H-K'$ color at late times indicates 
that the IR emission likely peaks in the mid-IR. To place a lower limit on 
the NIR luminosity we assume that the SED peaks in the $K'$ band. Following 
this assumption, the NIR excess constitutes a slowly varying luminosity 
of $\ga 2 \times 10^{8}$ \lsun\ for $>$ 1 yr.

The most likely explanation for this large luminosity in the NIR is warm 
dust. The observed $H - K'$ color from day 436 corresponds to a temperature 
of $\sim$1000 K, assuming a single-temperature blackbody. If the dust is 
radiating as a single blackbody with $T_{\rm peak} = 1000$ K, this would 
increase the above luminosity to $\ga 3 \times 10^{8}$ \lsun. 
We note that our NIR observations are only sensitive to the warmest dust; 
it is possible that cooler dust with $T \approx 
600$~K\footnote{\citet{smith08-2006gy} show that the combination of the 
peak luminosity and distance to the dust suggest an equilibrium temperature 
around $600$~K.} could dominate the IR 
emission, in which case the luminosity 
could be significantly higher than the values quoted above. 

The location 
of this dust, and whether it is newly formed or pre-existing at the time 
of the SN explosion, remain to be determined. The dust-cooling time is 
short, meaning that a prolonged heat source is needed to explain the 
extended excess. We consider four possibilities for heating the dust: (i) 
radioactive heating from $^{56}$Co decay, (ii) collisional excitation of 
pre-existing dust, (iii) heating via radiation from circumstellar interaction, 
and (iv) a late-time IR dust echo, where pre-existing dust is heated by 
the radiation produced while the SN was near its optical peak.

\subsubsection{Radioactive Heating from $^{56}$Co}

\begin{figure}
\epsscale{1.15}
\plotone{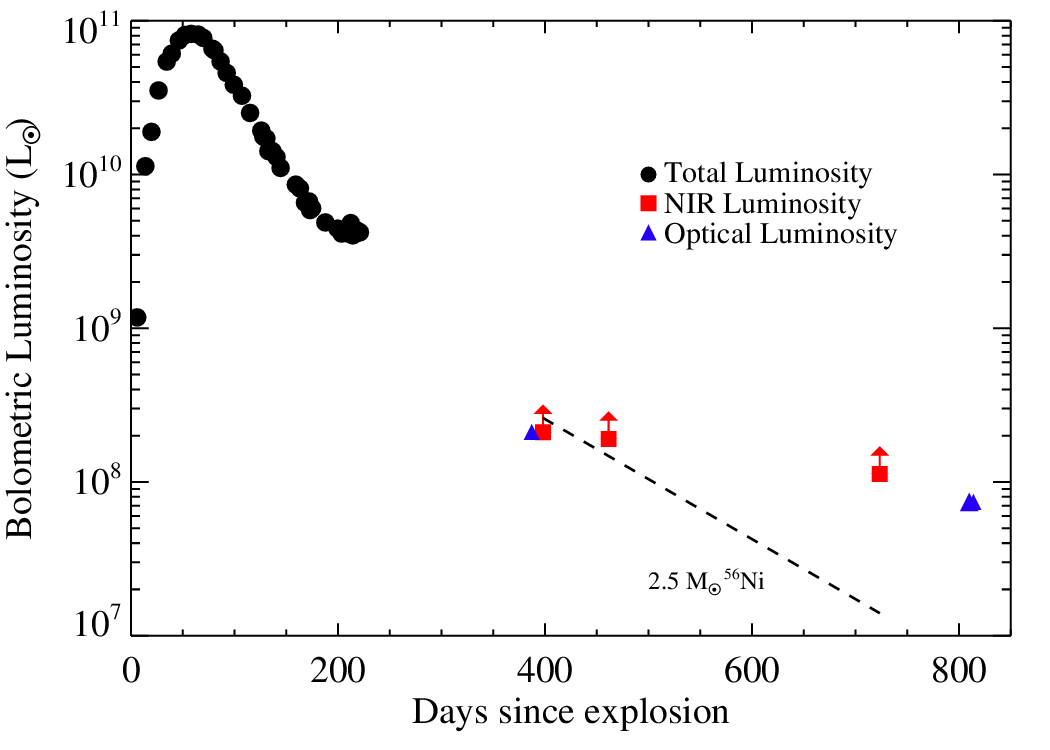}
\caption{Evolution of the bolometric luminosity of \sn\ during the first 800 
days after explosion. Optical data are shown as black circles, and NIR data 
are red squares. Early-time data ($<$ 250 days) are from \citet{smith10-2006gy}. 
The optical detection near day 400 is from \citet{kawabata09}. All other data 
are from this work. NIR measurements are only lower limits to the total IR 
luminosity, because our observations do not sample redward of 2.2 $\mu$m 
(see text). The dashed line 
shows the expected decline if the luminosity were powered by 
2.5 \msun\ of $^{56}$Ni. The flat nature of the light curve indicates 
that radioactivity is not the primary energy source for the late-time 
emission.} 
\label{bolo-lc}
\end{figure}

\citet{smith08-2006gy} noted that the observed $K'$-band decay between 2007 
September and 2007 December could be explained with radioactive heating 
from a minimum of 2.5 \msun\ of $^{56}$Ni, if a sufficient amount of 
dust formed in order to move the luminosity into the NIR (though they 
favored another interpretation; see below). 
We show the bolometric evolution of \sn\ through the first $\sim$800 days 
post explosion, including both optical and NIR detections, in 
Figure~\ref{bolo-lc}. From Equation~19 of \citet{nadyozhin94}, and the fact 
that the luminosity from $^{56}$Co decay dominates over $^{56}$Ni decay at 
times $\ga$ 2 weeks after explosion, we arrive at the following expression 
for the radioactivity-powered luminosity of a SN, assuming 100\% trapping of 
gamma-rays:

\begin{equation}
L_{^{56}{\rm Co}} = 1.45 \times 10^{43}\exp^{-t/(111.3~{\rm d})} 
M_{\rm Ni}/{\rm M}_\odot{\rm~erg~s}^{-1}, 
\label{Co-lum}
\end{equation}
where $t$ is the time since SN explosion in days, and $M_{\rm Ni}$ is the 
total mass of $^{56}$Ni produced. Using Equation~\ref{Co-lum} we also show 
in Figure~\ref{bolo-lc} the 
expected light curve from 2.5 \msun\ of $^{56}$Ni at times $>$ 400 days.
The early-time measurements ($<$ 250 days) come from \citet{smith10-2006gy}. 
The optical measurement near day 400 comes from photometric measurements by
\citet{kawabata09}, while the optical luminosity on day 810 comes from a 
direct integration of our {\it HST} observations (see $\S$ 3.3.1). The three 
late-time NIR luminosities represent lower limits based on the 
measured $K'$-band flux (see above, $\S$ 3.2). The decay rate of 
$^{56}$Co, 0.98 mag (100 day)$^{-1}$, is much
faster than the observed decay of the $K'$ flux from \sn, $\sim$0.2 mag 
(100 day)$^{-1}$. 
The late-time NIR excess declines at a rate that is too 
slow to be explained by $^{56}$Co heating alone.

The PISN model of \citet{nomoto07}, which provided 
good agreement with the early-time light curve of \sn\ after an artificial 
reduction of the total ejecta mass from their evolutionary calculation, was 
able to reproduce 
the late-time NIR luminosity observed by \citet{smith08-2006gy}. This model 
required less than 100\% efficiency in the conversion of gamma-rays (from
radioactive decay) to optical/NIR emission, which means that the light curve 
should decay {\it faster} than 0.98 mag (100 day)$^{-1}$. The possibility of 
a PISN was first invoked to explain the large peak luminosity 
of \sn\ \citep{ofek06gy, smith07-2006gy}. This scenario would have required 
the production of $\ga$ 10 \msun\ of $^{56}$Ni, which in turn would produce a 
large late-time luminosity that decays at the rate of $^{56}$Co. The late-time 
NIR luminosity is not accounted for in either the general PISN
models or the artificial model of \citet{nomoto07}; it therefore provides a 
serious challenge to the PISN hypothesis.\footnote{The pulsational 
pair-instability model of \citet{woosley07} has not, however, been excluded; 
see also \citet{smith10-2006gy}.} 

\subsubsection{Collisional Heating of the Dust}

Another possibility is that the dust existed prior to the SN explosion, at 
large distances from the explosion site, and was heated via collisions 
with the expanding material in the expanding blast wave. The intense 
UV/optical output from a SN at its peak vaporizes any dust in the 
vicinity of the SN \citep{dwek1983}. This radiation near peak creates a 
dust-free cavity into which the SN ejecta may expand at early times; however, 
the ejecta blast wave will eventually reach the edge of the dust-free 
cavity, at which point collisional excitations of the dust may generate NIR 
emission. 

The large peak luminosity of \sn, $8 \times 10^{10}$ \lsun\ 
\citep{smith10-2006gy}, provides significant challenges to this collisional 
heating scenario. \citet{dwek1983} shows that the radius of the dust-free 
cavity can be determined from the peak luminosity and the dust vaporization 
temperature, assuming the grain emissivity $Q_v$ is proportional to 
$(\lambda/\lambda_0)^{-1}$:

\begin{equation}
R_1({\rm pc}) = 23\left[\frac{\bar{Q}_\nu 
L_0({\rm L}_\odot)}{\lambda_0(\mu{\rm m}) T_v^5}\right]^{0.5},
\label{rad-eqn}
\end{equation}
where $R_1$ is the radius of the dust-free cavity, $\bar{Q}_\nu$ is the mean 
grain emissivity, $L_0$ is the peak luminosity, and $T_v$ is the dust 
vaporization 
temperature. Following \citet{dwek1983}, if we assume $\bar{Q}_\nu = 1$ and that 
$\lambda_0 = 0.2$~$\mu{\rm m}$, we find that $R_1 \approx 1.4 \times 10^{18}$ 
cm, for a vaporization temperature $T_v \approx 1000$~K. If the absorption 
coefficient is instead proportional to $\lambda^{-2}$, then the cavity becomes 
even larger. Based on the observed width of the H$\alpha$ line from \sn, 
\citet{smith07-2006gy} 
estimate the speed of the blast wave to be 4000 km s$^{-1}$, which would 
mean that this material would take $\sim$112 yr to reach the edge of the 
dust-free cavity. Even if there are ejecta traveling at more typical SN 
velocities of 10,000 km s$^{-1}$, it would take over 45 yr for this material to 
reach the edge of the dust-free cavity. Given that the NIR excess is present 
$\sim$1 yr following explosion, and that the SN ejecta could not reach 
the dust in such a short time, collisional heating of dust by SN ejecta 
cannot explain the observed NIR excess. Large grains will persist at smaller 
radii than smaller grains according to Equation~\ref{rad-eqn}, since the 
size of the dust-free cavity is $\propto \lambda_0^{-0.5}$, while the grain 
size, $a = \lambda_0/2\pi$. The existence of grains at $R_1 \approx V_{\rm SN} 
~t_{\rm SN} = 4000$ \kms\ $\times~1$ yr = 1.26 $\times 10^{16}$ cm, and thus 
possibly explain the NIR excess seen 1 yr after explosion, would require 
dust particles a factor $\sim 10^4$ larger than those assumed above, roughly 
corresponding to $\sim$1 mm size grains. Such large grains are rare and 
are inefficient radiators in the NIR \citep{spitzer}; therefore, they are 
unlikely to explain the observed NIR excess. 

\subsubsection{Newly Formed Dust}

The emission features that arise in SNe~IIn, as in \sn, originate from 
the interaction of the SN ejecta with a dense CSM. 
Shocked gas, located between the forward shock which is plowing into the CSM 
and a reverse shock of the SN ejecta, can cool, and if the density is large 
enough, it may form new dust grains. The cooling time for such dust is short, 
and can be estimated in the following manner: $\tau_{\rm cool} \approx E/L$ 
where $E$ 
is the thermal energy of a grain and $L$ is the grain luminosity. We estimate 
the thermal energy as $Nk_BT_{\rm dust}$, where $N$ is the total number of 
particles in a grain and $k_B$ is the Boltzmann constant. We assume the grains 
radiate as a blackbody: $L = 4\pi{a}^2 \sigma T^4_{\rm dust}$, where $a$ is 
the typical grain size and $\sigma$ is the Stefan-Boltzmann constant. For 
graphite grains of size $a \approx 1$ $\mu$m, and $\rho \approx 3$ g cm$^{-3}$, 
we have $N \approx 10^{12}$, meaning $\tau_{\rm cool} <$ 1 s. 
Furthermore, theoretical calculations by \citet{draine+li2001} show that the 
cooling time for PAH and silicate grains at $T \approx 1000$ K is $\la$ 10 s, 
for grains of all sizes. Nevertheless, 
a persistent heat source can generate an extended period of NIR emission. 
The physical conditions for post-shock dust formation are not necessarily 
easy to produce: strongly interacting SNe are often X-ray sources, which 
could prevent the formation of new dust grains in the post-shock gas. 
Evidence for the formation of dust in the post-shock region has been 
found for a few SNe, however, including SN 1998S \citep{pozzo2004}, 
SN 2006jc \citep{smith08-2006jc}, and SN 2005ip \citep{smith09-2005ip, fox2009}. 
This newly formed dust is then heated by the energy from the shock as it 
continues to propagate into the CSM, which gives rise to the prolonged NIR 
excess.

This scenario is difficult to reconcile with the case of SN 2006gy, however. 
Spectra taken around day 200 show a decline 
in the luminosity of the H$\alpha$ emission line, which indicates a 
reduction in the CSM interaction at this time \citep{smith10-2006gy}. 
Furthermore, as detailed by \citet{smith08-2006gy}, the lack of X-ray, 
H$\alpha$, and radio emission a little more than a year after the 
SN explosion implies that the observed NIR luminosity cannot be 
explained by shock-heated dust. In fact, the radio nondetection of \sn\ 
continues through day 638 \citep{2008ATel.1657....1B}. \citet{kawabata09} 
do claim the detection of very weak H$\alpha$ emission 
from \sn\ on day 394, but the inferred luminosity of $\sim$10$^{39}$ 
erg s$^{-1}$ is nearly three orders of magnitude lower than the observed NIR 
luminosity at this time. \citet{kawabata09} also claim an $R$-band detection 
of 19.4 mag on day 394 (though we note that \citealt{smith08-2006gy} report 
an upper limit of $R >$ 20 mag, and \citealt{agnoletto09} give an upper 
limit of $R >$ 20.3 mag, at a similar epoch), which would 
constitute an optical luminosity similar to that observed in the NIR. 
The source of this optical luminosity could potentially be the heat 
source for newly formed dust at day $\sim$400, though it cannot 
explain the similar NIR luminosity seen on day 723. While \sn\ was 
detected in the optical at $>$ 800 days to have a luminosity similar to 
that seen in the NIR, the very blue nature of optical data suggests that 
it is a scattered-light echo (see Section 3.3) of UV/optical emission from 
the SN peak. The light echo must originate {\it outside} the dust-forming 
region and therefore cannot heat the dust. While we cannot strongly rule 
out dust formation in the post-shock region of \sn, it is implausible 
to explain the late-time NIR luminosity via new dust, because there are 
no signs of a heating source that lasts for $>$ 1 yr.

\subsubsection{NIR Dust Echo}

Perhaps the most natural explanation for the late-time NIR excess is a 
dust-heated echo, as first discussed by \citet{smith08-2006gy}. In this 
scenario, the IR emission comes from dust near the explosion site which 
is heated by the early-time UV/optical emission from the SN. IR echo models 
by \citet{dwek1983} predict a fast rise (of order the rise time in the 
optical of the SN) 
in the IR, followed by an extended plateau (lasting $>$ 1 yr; see below) 
before the IR luminosity follows a rapid decline (where the IR flux 
declines as $e^{-t}/t$). The extended plataeu from an IR echo matches the 
qualitative behavior of \sn, which exhibits a very slow decline in the NIR: 
only a factor of $\sim$2 in flux over the course of $\sim$1 yr. 

\begin{figure}
\epsscale{0.8}
\plotone{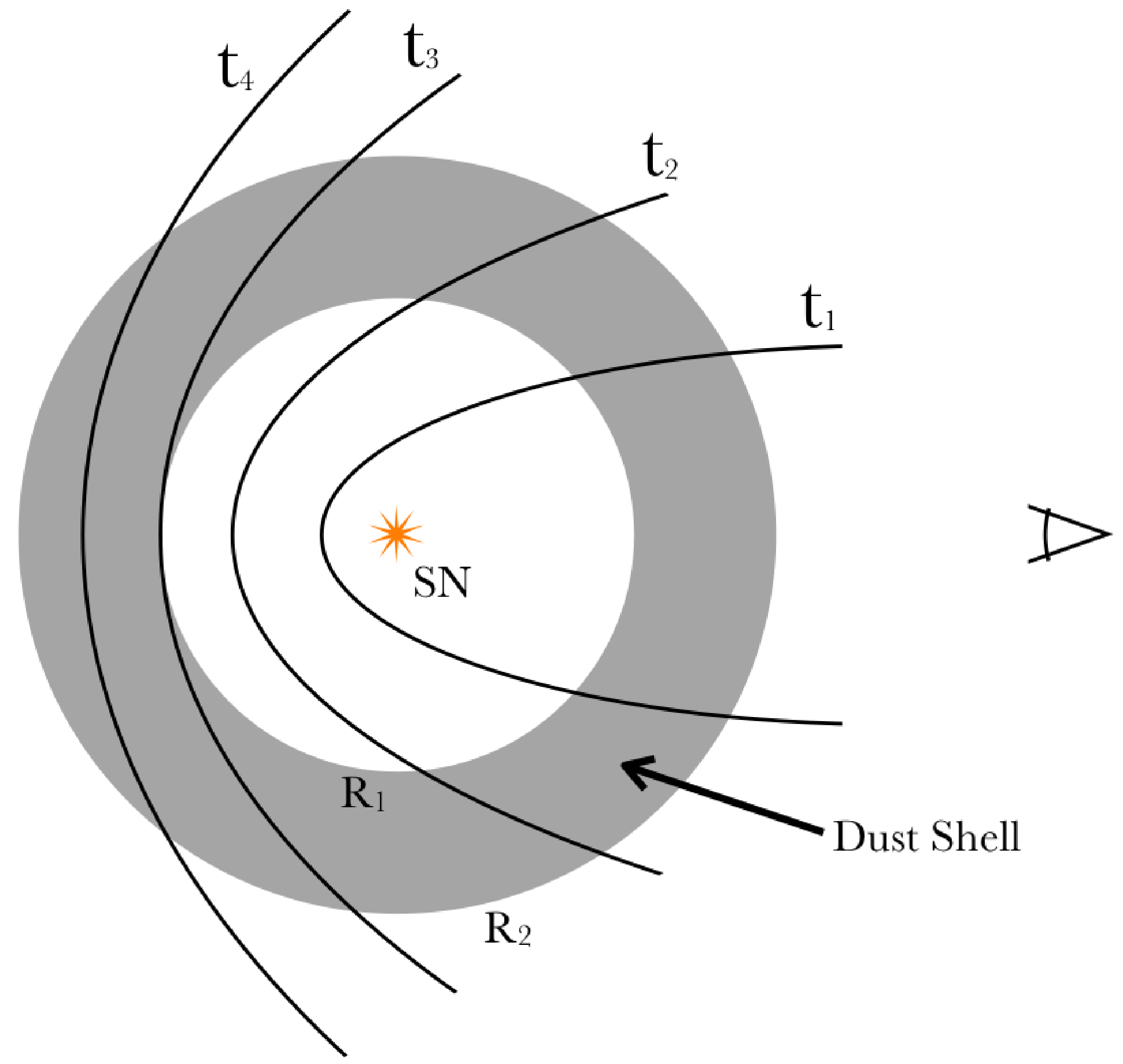}
\caption{Schematic diagram showing the evolution of an IR echo arising from a 
circumstellar dust shell, shown in the shaded area. The large luminosity 
from the SN vaporizes a dust-free 
cavity out to radius $R_1$. At early times the dust-emitting volume is 
small, shown as the area interior to the parabola marked $t_1$, as the 
reradiated light from other portions of the shell has not had sufficient 
time to reach the observer.  As time increases, so does the emitting volume, to 
the area within $t_2$, then $t_3$, etc. The echo luminosity is dominated by 
the warmest dust at $\sim$$R_1$. As the light parabola sweeps through the 
shell, new emission from dust at $\sim$$R_1$ continually becomes observable, 
leading to a plateau in the NIR light curve. The plateau continues  
until a time $t_3 \approx 2R_1/c$, at which point the parabola has swept past 
the inner edge of the shell. Additional emission continues to be observable 
($t_4$); however, this is from more distant, cooler dust and as a result 
the IR echo falls off the plateau and begins to decline. 
}
\label{dust-echo}
\end{figure}

Assuming isotropic optical emission from the SN, the plateau occurs because 
the hottest dust, which exists right at the 
edge of the dust-free cavity described above, dominates the emission. 
According to the view of a distant observer, the emitting volume is a series 
of paraboloid light fronts that expand throughout the dust-free cavity 
as shown in Figure~\ref{dust-echo} 
(see also Figure 1 of \citealt{dwek1983}). 
As a paraboloid expands it continually heats dust at $R_1$, which is what 
gives rise to the plateau in the light curve. After a time $\sim$2$R_1$/$c$, 
the vertex of a given paraboloid will reach the back edge of the cavity 
and will no longer be heating dust at ${R_1}$. Emission from the dust 
will be dominated by the UV/optical radiation produced by the SN at 
peak, and once this radiation sweeps past the back edge of the cavity 
the IR luminosity  begins to rapidly decline. Therefore, if we know 
the duration of the plateau phase we can determine the size of the 
dust-free cavity. Constraints on the cavity 
size can be determined by assuming that the echo starts at our first $K'$ 
observation on day 398 and ends on our last observation on day 723. In this 
case we find that $R_1 \approx 4 \times 10^{17}$ cm; however, it would be 
contrived if our 
observations were to perfectly bookend the plateau phase of the echo: the 
echo almost certainly started prior to day 398 and continued beyond day 723. 
In fact, the model predicts that the echo should start well before day 
$\sim$400. There is mild evidence for a NIR echo as early as day $\sim$130: 
the $H$-band and $K_s$-band flux is (respectively) $\sim 2\sigma$ and 
$\sim1\sigma$ greater than their expected values based on an extrapolation 
of optical spectra (see Figure~\ref{early-lc}). If we instead assume that the slight NIR excess seen at day $\sim$130 (see 
Figure~\ref{early-lc}) is the rise of an IR echo, then we find that 
$R_1 \approx 8 \times 10^{17}$ cm, which shows reasonable agreement with the value 
we calculated for the dust-free cavity in $\S$ 3.2.2, 
$R_1 \approx 1.4 \times 10^{18}$ cm, considering the uncertainties in the 
dust temperature and the likelihood that the plateau extends beyond day 723.

Our NIR observations are only sensitive to the warmest dust, but if we assume 
that the NIR luminosity peaks in the $K'$ band, then a lower limit to the 
total energy emitted in the IR is $E_{\rm IR} \approx 2 \times 10^{8}$ \lsun\ 
$\times~{600}$ days, or $\ga$ 4$\times 10^{49}$ erg. This is comparable to 
the canonical optical output of a normal SN~II. Typically, this would 
pose a significant problem for the IR echo hypothesis, as it would require 
of order unity of the original optical radiation from the SN to be absorbed 
and reradiated in the NIR. Indeed, this is precisely the argument used by 
\citet{fox2009} to argue against the light-echo hypothesis for SN 2005ip. 
In the case of \sn\ this is not a problem, however, because the NIR 
emission falls well within the available energy budget to heat the dust. 
After applying a bolometric correction, \citet{smith10-2006gy} find that 
$\sim 2.5 \times 10^{51}$ erg were emitted by \sn\ during the first 
$\sim$220 days of the light-curve evolution, which means that less than 
2\% of this energy has been absorbed and reradiated by the dust. 
This value is comparable to those found by \citet{dwek1983} for SNe 1979C 
and 1980K.

If we assume that the dust radiates as a blackbody, 
we can estimate the optical depth of the dust shell as $\tau 
\approx E_{\rm IR}/(E + E_{\rm IR}) \approx 0.016$. We note that like 
$E_{\rm IR}$ above, this value constitutes a lower 
limit for $\tau$. Even if the NIR luminosity is a factor of 10 larger than 
our lower limits, this would imply $\tau \approx 0.1$, which corresponds 
to an optical extinction of $A_V = 1.086 \tau \approx 0.1$ mag. This is far 
less than the total observed extinction from the host, $A_{V, {\rm host}} 
\approx 1.67$ mag \citep{smith07-2006gy}, suggesting that the dusty shell 
responsible for the IR echo cannot produce all of the non-Galactic reddening 
observed toward \sn. That there is a considerable amount of dust along the 
line of sight, which is not in the CSM of \sn, should not come as a 
surprise given that there is a prominent dust lane in NGC 1260 (see 
Figure~\ref{false-color}).

With this information in hand, it is now possible to estimate the total 
dust mass of the shell that creates the NIR echo. According to 
\citet{dwek1983} for a dust shell with density $n_{\rm d} \propto r^{-2}$, 

\begin{equation}
M_{\rm d} = 4\pi\left(\frac{4\rho_{\rm gr}a}{3\bar{Q}_\nu}\right)\tau_{\rm d} 
\times \frac{R_1R_2^2}{R_2-R_1},
\label{dust-mass}
\end{equation}
where $M_{\rm d}$ is the total dust mass in the shell, $\rho_{\rm gr}$ is the 
grain density, $a$ is the grain radius, $\tau_{\rm d}$ is the optical depth, 
$\bar{Q}_\nu$ is the mean 
absorption efficiency, and $R_1$ and $R_2$ are the inner and outer radii 
of the dust shell, respectively. Above we estimate the inner radius of the 
dust shell to be 
$\sim$8$\times 10^{17}$ cm. The value of the outer radius can be deciphered 
from the decay of the NIR light curve following the plateau. As we have not 
yet observed the NIR decay, we adopt an outer radius of 2$R_1$, which 
provides a {\it lower limit} to the dust mass in the shell. Finally, we follow 
the same assumptions as \citet{dwek1983} about the dust properties: 
$\rho_{gr}$ = 3 g cm$^{-3}$, $a = 0.1\, \mu$m, and $\bar{Q}_\nu$ = 1. 
With these asumptions we find a 
total dust mass of $M_d \approx 0.1$ \msun. Assuming a typical value for the 
dust-to-gas ratio, 1:100, the total mass of the circumstellar 
shell is $\sim$10 \msun. This value shows good agreement with the initial 
estimate of 5--10 \msun\ from \citet{smith08-2006gy}, which was based
only on observations taken between 400 and 461 days after explosion. We note 
that  $\sim$10 \msun\ constitutes a lower 
limit to the total mass of the shell, because the actual value of the 
duration of the plateau and $E_{\rm IR}$ could potentially be much larger 
than the values we adopted above.

At a distance of $\sim$10$^{18}$ cm from the SN explosion site, it is not 
immediately evident whether the dust giving rise to the NIR echo is part of 
the progenitor's CSM or the local interstellar medium (ISM). 
\citet{smith08-2006gy} argue that 
the large dust mass needed at $\sim$10$^{18}$ cm could be explained if 
the progenitor passed through a phase of eruptive mass loss $\sim$1000--1500 
yr prior to explosion, in an event similar to observed outbursts 
from $\eta$ Car. Pre-existing dust in the walls of a giant H~II 
region, where a massive star like the progenitor of SN 2006gy
might have lived, cannot account for the IR echo because at
typical distances of $\sim$10 pc or more from the SN, the dust would be 
far too cold to reproduce the NIR excess.

\subsection{Observed Color Evolution from HST Observations}

\citet{chevalier1986} predicts that any SN with an IR echo due to 
pre-existing CSM dust should also show a faint scattered-light 
echo in the optical as well. He shows that this dust 
can lead to optical emission characterized by a $\lambda^{-\alpha}$ scattering 
law, where $\alpha$ is some value between 1 and 2. More than a year after 
explosion \sn\ had faded 
rapidly in the optical \citep{smith08-2006gy,agnoletto09,kawabata09}. 
Therefore, we obtained {\it HST} images to search for a possible late-time 
scattered-light echo.

Our optical detection on day 810 shows a significant change in the optical 
decline rate of \sn. During the interval 200--400 days the SN faded by $\ga$ 3 
mag in the optical \citep{smith08-2006gy}, whereas the decline over 
days 400--810 was only $\sim$1 mag. In addition to this reduced rate of 
decline, the SN underwent significant color evolution. Assuming 
that $F555W$ and $F814W$ approximate the $V$ and $I$ bands, respectively,
we find that $V-I = 0.60$ mag on day 810. By contrast, with a 
12,000~K color temperature at peak \citep{smith10-2006gy}, the 
observed color of \sn\ would have been $V-I \approx 1.01$ mag. Thus, 
the spectrum on day 810 is 
{\it bluer} than the spectrum at peak. Indeed, the $\sim$0.4 mag color 
change shows excellent agreement with a $\lambda^{-1}$ scattering law. 
A $\lambda^{-2}$ scattering law, on the other hand, would predict a $-0.83$ 
mag change to the $V-I$ color, or roughly $V-I \approx 0.20$ mag during the 
late-time epochs.

This behavior is similar to what one would expect from a scattered-light 
echo: as in the case of an IR echo, when successive paraboloids sweep out 
to progressively larger radii, dust in the circumstellar (or in some cases the 
interstellar) environment can scatter that light toward the observer, 
creating a plateau in the optical light curve of the SN. At the same time, 
scattering preferentially selects shorter wavelengths, which results in a 
spectrum that is bluer than that of the SN at peak. This is the precise 
behavior observed for the Type Ia SNe 1991T \citep{schmidt94}, 1998bu 
\citep{cappellaro2001}, and 2006X \citep{wang2008}, which all showed 
significant departures from the expected decline rate of SNe~Ia. 
Spectra taken roughly two years after maximum, for SNe 1991T and 1998bu, and 
$\sim$10 months after maximum for SN 2006X, show emission features similar 
to those seen in the spectra near peak superposed on top of a blue 
continuum for these SNe. Light echoes have also been observed around 
a number of core-collapse SNe, such as SN 2003gd \citep{sugerman2005, 
vandyk2006}, SN 1993J \citep{sugerman+crotts2002}, and SN~1987A 
(see \citealt{sugerman87A-05}, and references therein).

\begin{figure}
\epsscale{1.15}
\plotone{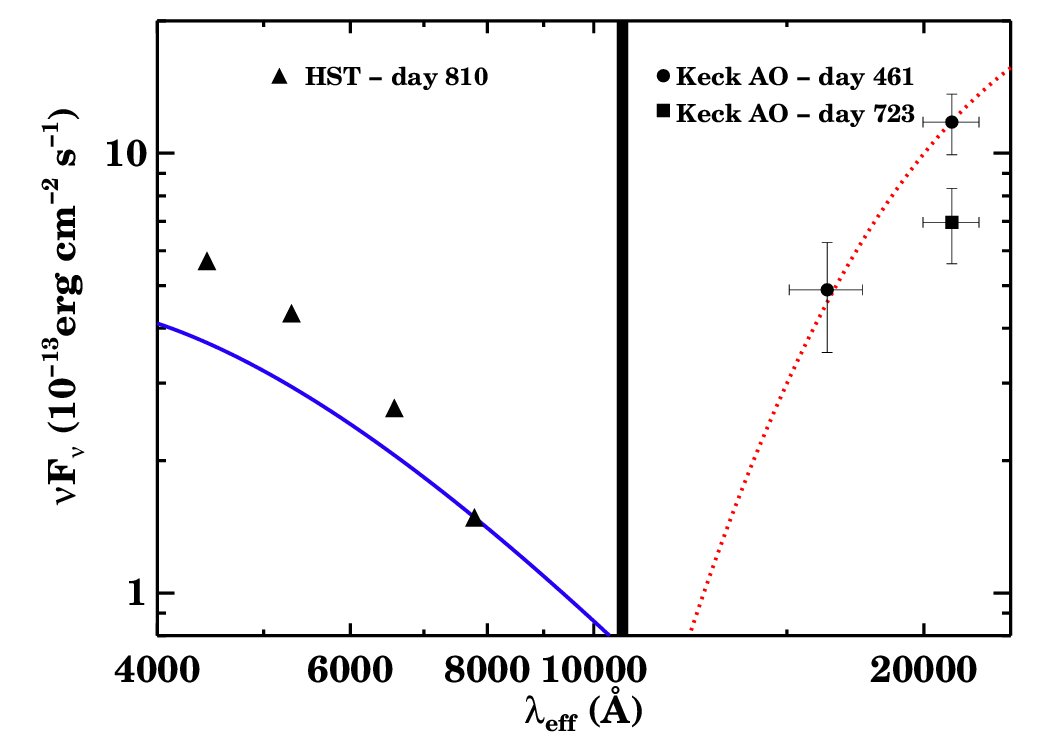}
\caption{Late-time SED of \sn\ showing evidence for two distinct emission 
components. We show the flux in the optical ({\it HST}, 
day 810) and NIR (Keck AO, days 461 and 723), corrected for host-galaxy 
and Milky Way reddening. The vertical solid line is to remind the reader 
that optical and NIR observations were not taken simultaneously. The solid 
blue line shows a 12,000~K blackbody, normalized to the $F814W$ detection. 
The red dotted line shows a single-component blackbody fit ($T = 1000$ K) 
to the $H$ and $K'$-band 
detections from day 461 (note that this fit has zero degrees of freedom). The 
IR evolution is slow, and the detection from day 723 indicates the degree 
of fading in the NIR. The SED shows a rapid decline toward the red portion 
of the optical, as well as a strong rise toward the red within the NIR. 
These characteristics are precisely those predicted by the IR and 
scattered-light echo models.} 
\label{late-sed}
\end{figure}

In Figure~\ref{late-sed} we show the late-time SED of \sn, including our NIR 
detections on days 461 and 723 and the optical detections from day 810. The 
solid blue line represents a single-component blackbody at $T = 12,000$ K, roughly 
the peak temperature of \sn\ \citep{smith10-2006gy}, normalized to our 
$F814W$ detection. The SED clearly shows that the late-time emission is 
significantly bluer than a blackbody spectrum. We note that the early-time 
spectra of \sn\ show considerable 
line blanketing blueward of $\sim$4000 \AA, such that the emission 
decreases blueward of our $F450W$ detection. The NIR emission rapidly 
rises toward the IR, suggesting that there is considerable IR 
emission to which our NIR measurements are not sensitive.

\begin{figure}
\epsscale{1.15}
\plotone{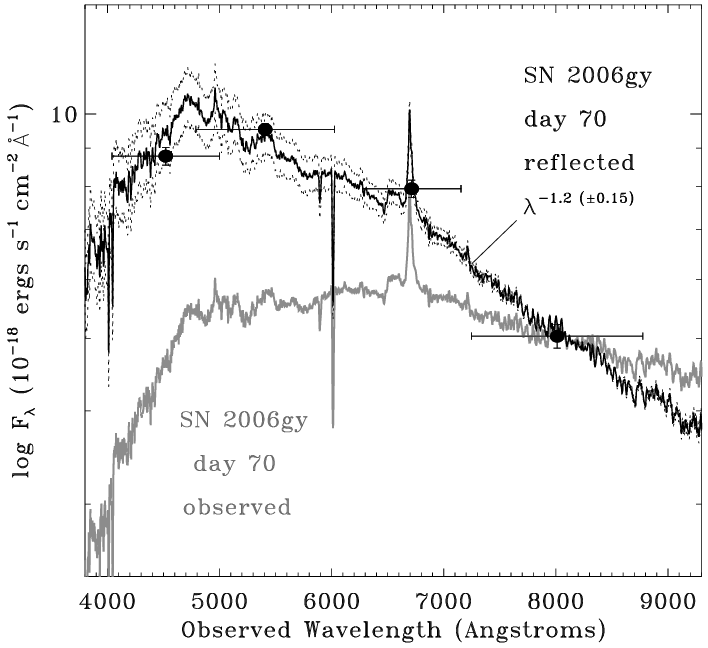}
\caption{Evidence for a scattered-light echo from the blue optical emission 
from SN~2006gy. The {\it HST}/WFPC2 photometry of \sn\ (black dots) is
compared to observed and reflected spectra, both normalized to the 
{\it HST} $F814W$ flux density. The gray line is the observed spectrum of 
SN~2006gy on day 70 during its peak luminosity phase, from 
\citet{smith10-2006gy}. We plot the observed wavelength (i.e., not corrected 
for the SN redshift), and the spectrum has no correction for reddening. 
The black line (and dotted lines above and below) show the same spectrum, 
but adopting an assumed wavelength dependence for the dust reflectance 
proportional to $\lambda^{-1.2(\pm0.15)}$.} 
\label{reflect-spec}
\end{figure}

Figure~\ref{reflect-spec} illustrates the observed {\it HST} photometry on 
day 810, as well as the observed spectrum (grey solid line) of \sn\ 
obtained near peak luminosity \citep{smith10-2006gy}. The observed late-time 
photometry 
is clearly bluer than the spectrum at the time of peak luminosity (note that 
no correction for reddening has been made to either the observed spectrum or 
the late-time photometry). The blue colors at late times are difficult to 
explain with ongoing CSM interaction because SNe IIn typically exhibit 
relatively constant temperatures of $\sim$6000--6500 K at late times (e.g., 
\citealt{smith10-2006gy}), which would be redder than the spectrum at peak 
which had a temperature of $\sim$12,000 K. A bluer color would be expected, 
however, if the optical emission at the time of peak luminosity is reflected 
by dust grains with a size smaller than the observed wavelength, which leads 
to the spectrum characterized by a $\lambda^{-\alpha}$ scattering law, as 
mentioned above. 
Figure~\ref{reflect-spec} shows that the observed spectrum from day 71 can be 
modified by a scattering law with $\alpha = 1.2 \pm 0.15$ to adequately model 
the late-time {\it HST} photometry. We do not have a late-time spectrum to 
confirm that the observed line strengths are consistent with the reflected 
peak luminosity, and so this precludes a more detailed model of the 
scattering dust at this time. 

\subsection{Location of the Scattering Dust}

The location of the scattering dust is of interest as it can help 
determine properties of the environment local to \sn. In Section~3.2.4  
we show that the CSM dust responsible for the NIR echo cannot account for 
all the reddening observed toward \sn, suggesting that the SN light 
passes through external, non-CSM dusty regions in the host galaxy prior 
to reaching us. Therefore, the scattered-light echo could originate in 
a number of different locations.

With only a single epoch of the unresolved scattered-light echo, a 
precise determination of the location of the scattering dust cannot be 
obtained. Nevertheless, we can gain some insight into the dust location 
based on the 
total optical and NIR emission. For the case where the NIR-echo dust also 
gives rise to the scattered-light echo, and assuming the SN can be modelled 
as a single short pulse of UV/optical radiation, the ratio of the 
scattered-light luminosity ($L_{\rm s}$) to the IR luminosity at 
late times can be shown to be \citep{chevalier1986}

\begin{eqnarray}
\frac{L_{\rm s}}{L_{\rm IR}} & = & \frac{\omega}{1-\omega} 
\left(1-\frac{(ct)^2}{4R_2^2}\right) \frac{(1-g)^2}{2g^2}  \label{eqn2} \\
& & \times \left\{\frac{(1-g)^2+2g}{[(1-g)^2+4g]^{1/2}}-\frac{(1-g)^2+gct/R_2}{[(1-g)^2+2gct/R_2]^{1/2}}\right\} \nonumber,
\end{eqnarray}
where $\omega$ is the dust albedo, $c$ is the speed of light, $t$ is the time 
since the SN explosion, $R_2$ is the outer radius of the dust shell, and $g$ 
is a measure of the degree of forward scattering. Once again, as in 
Section~3.2.4 we have assumed a dust density $n_{\rm d} \propto r^{-2}$. 
The case $g$ = 0 corresponds to isotropic scattering and $g$ = 1 is 
completely forward scattering. Note that in Equation~\ref{eqn2} we have 
assumed that $ct/2 < R_1$, where $R_1$ 
is the inner radius of the dust shell, and that the albedo is independent 
of frequency. Empirical estimates and numerical calculations \citep{white1979} 
show $g \approx$ 0.6, which we adopt here. Models and observations of 
interstellar dust show that for near-UV, optical, and NIR emission, which 
dominate the output from \sn\ near peak, the dust 
scatters light at these wavelengths with $0.4 \la g \la 0.7$ (see 
\citealt{draine2003}, and references therein). For the dust albedo we 
adopt $\omega \approx$ 0.6 \citep{mrn1977}. For the albedo, models and 
observations typically constrain $0.5 < \omega < 0.7$, for scattered near-UV, 
optical, and NIR light \citep{draine2003}. As mentioned above, we cannot 
constrain $R_2$ with our 
current observations, but we do know that $R_2$ must be greater than 
$R_1$, and the value of Equation~\ref{eqn2} does not change when 
$R_2 \rightarrow \infty$. For reasonable limits to the outer radius, 
$1.5R_1 < R_2 < \infty$, we find that $L_{\rm s}/L_{\rm IR} \approx 0.88$--0.60. 

Direct integration of the {\it HST} optical flux, which extends from 
$\sim$4500--8000 \AA, yields an optical luminosity of 
$\sim$(7.4$\pm$0.5)$\times 10^{7}$ \lsun. This value is somewhat 
uncertain because our data do not extend into the UV; however, 
optical spectra taken near peak show that line blanketing severely reduces 
the flux blueward of $\sim$4000~\AA\ \citep{smith10-2006gy}, so this 
uncertainty should not have a significant effect on the total scattered-light 
luminosity. If we assume that the NIR flux continues to decline at the same 
rate as that observed between days 398 and 723, then assuming that the NIR 
emission peaks in the $K'$ band (see Section 3.2), we find a 
{\rm lower limit} to the IR luminosity (again, because our NIR observations 
only probe emission from the hottest dust) of $L_{\rm IR} > 8.6 \times 10^{7}$ 
\lsun\ on day 810. Therefore, on day 810, $L_{\rm s}/L_{\rm IR} < 0.86$, 
consistent with the predicted values above in the case where the scattering 
dust is the same dust responsible for the IR echo.

\begin{figure}
\epsscale{1.0}
\plotone{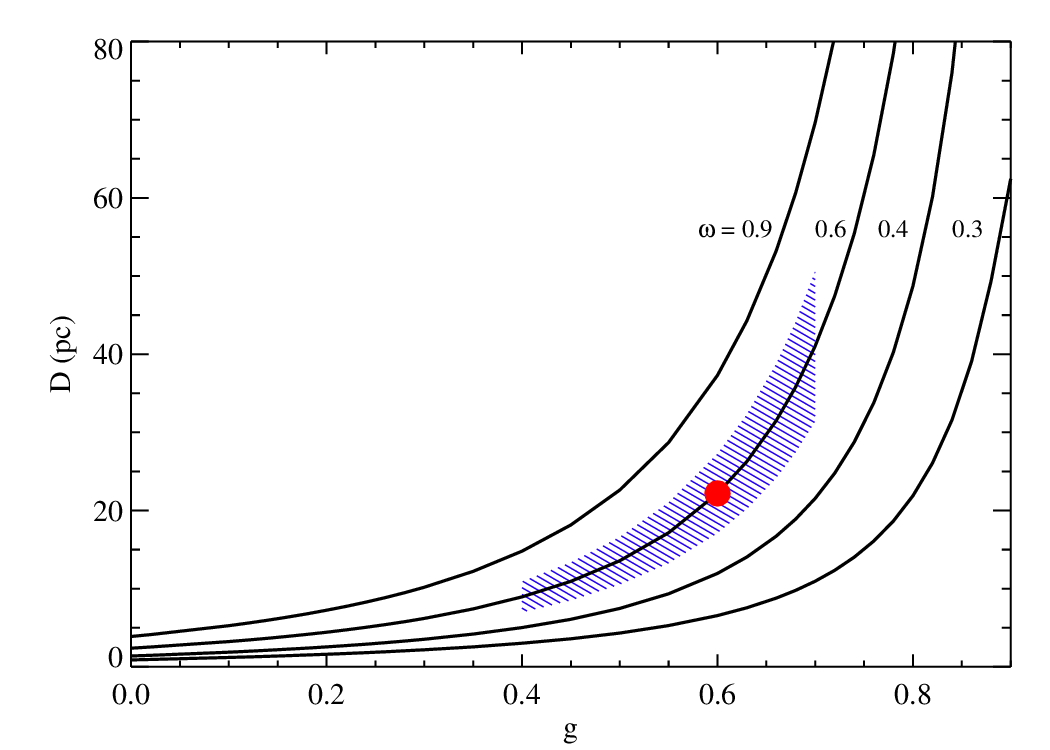}
\caption{Location of the scattering dust, $D$, as a function of the degree of 
forward scattering, $g$, for various values of the albedo, $\omega$. The large 
red point highlights our adopted values of $g = 0.6$ and $\omega = 0.6$, 
which corresponds to $D \approx 20$ pc. The shaded area encompasses the full 
range of values expected for the scattering of near-UV, optical, and NIR light 
by interstellar dust, $0.4 < g < 0.7$ and $0.5 < \omega < 0.7$ (see text). 
The shaded area corresponds to $D \approx 10$--40. Note that this figure only 
applies to the case where the scattering grains are external, and unrelated, 
to the dust giving rise to the NIR echo.} 
\label{dust-dist}
\end{figure}

The alternative possibility is that the scattering dust is not located in 
the CSM, instead existing at some other location in NGC 1260 along the 
line of sight. \citet{cappellaro2001} inferred that this was the case 
with SNe 1991T and 1998bu, and like those two SNe we know that \sn\ has 
a significant amount of dust located along the line of sight. As shown by 
\citet{cappellaro2001}, if we approximate the SN light curve as a short 
pulse with duration $\Delta{t_{{\rm SN}}}$, then the scattered-light 
luminosity is 
\begin{equation}
L_{\rm echo}(t) = L_{{\rm SN}} \Delta{t_{{\rm SN}}} f(t),
\end{equation}
where $L_{{\rm SN}} \Delta{t_{{\rm SN}}}$ can be obtained via direct 
integration of the observed light curve, 
$L_{{\rm SN}} \Delta{t_{{\rm SN}}} = \int_0^{+\infty} L_{{\rm SN}}(t)dt$, 
and $f(t)$ is the fraction of light scattered to the observer. Under the 
assumption that light is being scattered by dust in a sheet of 
thickness $\Delta{D} \ll D$, where $D$ is the distance between the dust 
and the SN, then \citep{chevalier1986,cappellaro2001}

\begin{equation}
f(t) = \frac{c}{8\pi}\frac{\omega\tau}{D+ct}\frac{1-g^2}{\left\{1+g^2-2g[D/(D+ct)]\right\}^{3/2}},
\label{eqn4}
\end{equation}
where $\tau$ is the optical depth of the dust. Using the relation 
$A_V = 1.086 \tau_V$, and the fact that $A_V$ outside the 
dusty CSM is $\sim$1.6 mag, we find $\tau \approx 1.5$. From 
\citet{smith10-2006gy},
we know that $L_{{\rm SN}} \Delta{t_{{\rm SN}}} \approx 2.5 \times 10^{51}$ erg. 
Therefore, using $L_{\rm echo} \approx 2.9 \times 10^{41}$ erg s$^{-1}$, 
from the direct integration of the optical SED, we 
find (from Equation~\ref{eqn4}) that dust located at $D \approx 20$ pc can 
explain the observed scattered-light luminosity. This value for $D$ is 
dependent upon our assumptions about the albedo and the degree of forward 
scattering, $g$. To show how $D$ 
changes based on these assumptions, we show $D$ as a function of $g$ for 
fixed values of the albedo $\omega =$ 0.3, 0.4, 0.6, and 0.9 in 
Figure~\ref{dust-dist}. The large red point shows the solution for our 
assumed values of $g = 0.6$ and $\omega = 0.6$. We also shade the area 
corresponding to the expected range of values for $g$ and $\omega$ mentioned 
above. This shaded area corresponds to $D \approx 10$--40 pc, as the location 
of the scattering dust. 

Our observations appear to be consistent with one of two scenarios: either 
(i) the scattered-light echo is due to CSM dust, which has also been 
heated and is radiating in the IR, or (ii) a dusty region $\sim$10--40 pc from 
the SN is responsible for the late-time optical luminosity. This latter 
scenario seems very plausible if the progenitor of \sn\ was a very 
massive star, $\ga$ 100 \msun\ \citep{smith10-2006gy}. If this were the 
case, we might expect that the progenitor resides in the center of a 
giant H~II region with multiple dense, dusty clouds nearby. This scenario 
might be very similar to the Carina nebula surrounding $\eta$ Car, 
which has multiple dusty molecular clouds only 10--20 pc from $\eta$ 
Car (e.g., \citealt{smith+brooks2007}). 
If the progenitor of \sn\ were surrounded by such a nebula, then our 
assumption of a single, thin scattering surface would no longer be 
valid, and the scattering dust could potentially be located at a number 
of different locations.

With only a 
single epoch of observations, we cannot distinguish between the two 
possible dust locations. One additional epoch of optical imaging should prove 
sufficient to determine which scenario is correct. As long as $D \gg ct$, 
then $f(t)$ is roughly constant, whereas the models of \citet{chevalier1986} 
show a continuous decline in the scattered-light echoes when $g$ = 0.6 
and the scattering dust is in the CSM. A relatively flat optical light 
curve would put the dust at $\sim$20 pc, whereas significant decline 
from the observed day 810 flux would suggest CSM dust. Finally, we note that 
the prospects for fully resolving the \sn\ light echo are dim: at a distance 
of 73 Mpc it would take several decades before the light echo could be 
resolved with high spatial resolution images (i.e., comparable to {\it HST} 
in the optical).

\section{Conclusions and Future Work}

We have presented new observations of \sn, including early-time NIR data from 
PAIRITEL, as well as optical and NIR detections more than two years after 
explosion. These new data, combined with other late-time observations,
provide evidence for IR and scattered-light echoes, as follows.
\begin{enumerate}

\item There is no radio or X-ray counterpart to SN~2006gy, suggesting that 
CSM interaction is weak or nonexistent at late times. 
\item There is (statistically weak) evidence for a growing NIR excess around 
day $\sim$100. 
\item The decline of both the optical and NIR emission is slower than that 
of $^{56}$Co, which rules out a pair-instability SN as the cause of the 
extreme peak optical luminosity for \sn.
\item Emission from warm dust explains the red NIR color, while an IR 
echo is needed to explain the long-lived NIR emission. 
\item The late-time optical emission has a bluer spectrum than the SN at 
peak optical emission, which can be explained with a scattered-light echo.

\end{enumerate}

Given our interpretation for the late-time optical and NIR emission arising 
from a dust shell at $\sim$8$\times 10^{17}$ cm from the SN, what might we 
expect from future observations of \sn? The IR-echo models of 
\citet{dwek1983} show that after a time $t \approx 2R_1/c$, the IR 
luminosity falls 
off the plateau and exhibits significant decline. Based on our predicted value 
for the size of the dust-free cavity (see $\S$3.2.2), we would expect 
that the plateau phase of the IR echo has ended, and that the IR flux is now 
in steady decline. Future observations of this decline will place limits on 
the outer extent of the dust shell, which will in turn provide better limits 
on the total mass of the dust shell. Our ability to predict 
the behavior of the scattered-light echo is limited by the current degeneracy 
concerning the physical location of the scattering dust. If the 
same dust is responsible for both the IR and scattered-light echoes, then we 
would expect that the optical emission is now fading, in a manner similar 
to what ought to be seen in the NIR. Our observations are also consistent 
with dust located at $\sim$10--40 pc from the SN, in which case we might 
expect the scattered-light echo to continue at roughly a 
constant flux for several more years. Continued observations with 
{\it HST} would allow us to distinguish between these two situations.

The shell-shock model of
\citet{smith+mccray07} may also explain the peak 
luminosity of several other VLSNe: SNe 2005ap, 2006tf, 2008es 
\citep{quimby05ap, smith06tf, miller08es, smith10-2006gy}. 
This model requires large CSM 
densities at a distance of a few $\times~10^{15}$ cm, which can be 
accomplished if the SN progenitor undergoes significant ($\sim$0.1--1 \msun\ 
yr$^{-1}$) mass loss during the few decades prior to explosion. The most 
luminous LBVs have observed 
mass shells with $\ga$10 \msun\ (\citealt{smith+owocki2006}, and references 
therein), indicative of giant eruptions 
\citep{humphreys1994} but not LBVs in their more typical state 
with less violent wind variability (see \citealt{svdk2004} for a general 
reference on LBV winds). The possible connection between large CSM densities 
and VLSNe may suggest that their progenitors are 
LBVs \citep{smith10-2006gy}. Furthermore, the direct identification of the 
progenitor of the Type IIn SN~2005gl showed it to be an LBV \citep{galyam07, 
galyam09}, 
which strengthens the connection between SNe~IIn and LBV progenitors. 
For the case of \sn, \citet{smith08-2006gy} 
argue that the dusty shell at $\sim$10$^{18}$ cm could exist if the 
progenitor underwent an LBV-like eruptive mass-loss phase 
$\sim$1200 yr prior to the SN. IR echoes for other VLSNe have not been 
reported to date, but we strongly encourage a search for them. 
If a substantial fraction of them exhibit late-time characteristics similar 
to \sn, this may be 
suggestive of a common timescale, $\sim$1000 yr and $\sim$10 yr, 
for extreme mass loss in the progenitors of VLSNe. The combination of a 
VLSNe and IR echo would point to multiple phases of eruptive mass loss, 
which is reminiscent of LBV behavior. The instability driving these 
eruptions still remains unclear (one possibility is the pulsation 
pair instability described by \citealt{woosley07} to explain \sn), and 
we encourage future observational and theoretical work to better 
characterize these systems.

\acknowledgments

We thank Schuyler Van Dyk for his assistance in the preparation of the 
{\it HST} snapshot proposal. Andy Becker, Maryam Modjaz, Dovi Poznanski,
and the anonymous referee provided many useful suggestions that helped
to improve this paper. We thank Cullen Blake, Dan Starr, and Emilio 
Falco for their assistance in the operation of PAIRITEL.

This publication makes use of data products from the Two Micron All
Sky Survey, which is a joint project of the University of
Massachusetts and the Infrared Processing and Analysis
Center/California Institute of Technology, funded by the National
Aeronautics and Space Administration (NASA) and the National Science
Foundation (NSF).

A.A.M. is supported by the NSF Graduate Research Fellowship Program. 
J.S.B.'s group is partially supported by NASA/{\it Swift} grant \#NNG05GF55G 
and a Hellman Faculty Award. A.V.F.'s SN group at UC Berkeley is grateful
for support from NSF grants AST-0607485 and AST-0908886, the TABASGO 
Foundation, and NASA grants AR-11248 and GO-10877 from the Space
Telescope Science Institute, which is operated by AURA, Inc., under
NASA contract NAS 5-26555. J.X.P. is partially supported by 
NASA/{\it Swift} grant NNX07AE94G and NSF CAREER grant AST-0548180.
The Peters Automated Infrared
Imaging Telescope is operated by the Smithsonian Astrophysical
Observatory (SAO) and was made possible by a grant from the Harvard
University Milton Fund, the camera loan from the University of
Virginia, and the continued support of the SAO and UC Berkeley. The
PAIRITEL project is partially supported by NASA/{\it Swift} Guest
Investigator Grant \#NNG06GH50G.
KAIT and its ongoing operation were made
possible by donations from Sun Microsystems, Inc., the Hewlett-Packard
Company, AutoScope Corporation, Lick Observatory, the NSF, the
University of California, the Sylvia \& Jim Katzman Foundation, and
the TABASGO Foundation.
Some of the data presented herein were
obtained at the W. M. Keck Observatory, which is operated as a
scientific partnership among the California Institute of Technology,
the University of California, and NASA; it was made possible by the generous
financial support of the W. M. Keck Foundation. The authors wish to
recognize and acknowledge the very significant cultural role and
reverence that the summit of Mauna Kea has always had within the
indigenous Hawaiian community; we are most fortunate to have the
opportunity to conduct observations from this mountain.



\end{document}